\documentclass[journal]{IEEEtran}
\usepackage{amsmath,amsfonts}
\usepackage{subcaption}

\usepackage[algo2e,ruled,vlined,linesnumbered]{algorithm2e}
\SetInd{0em}{0.5em}
\SetKwComment{Comment}{/\!\!/}{}

\usepackage{multirow}
\usepackage{array}
\usepackage{textcomp}
\usepackage{stfloats}
\usepackage{url}
\usepackage{verbatim}
\usepackage{graphicx}
\usepackage{cite}
\usepackage{subcaption}
\usepackage{hyperref}
\hypersetup{%
    colorlinks = false,
    hidelinks
    }

\usepackage{pgf}
\usepackage{pgfplots}
\usepackage{tikz}
\usetikzlibrary{spy}
\pgfplotsset{compat=1.18}

\usepackage{svg}
\usepackage{epstopdf}
\usepackage{physics}
\usepackage{makecell}

\usepackage{array}
\newcolumntype{H}{>{\setbox0=\hbox\bgroup}c<{\egroup}@{}}
\usepackage[normalem]{ulem}

\usepackage{format} %

\definecolor{burgundy}{rgb}{0.5, 0.0, 0.13}

\begin{document}
\title{Noisy MRI Reconstruction via MAP Estimation with an Implicit Deep-Denoiser Prior}

\author{%
Nikola Janju\v{s}evi\'{c}$^1$, 
Amirhossein Khalilian-Gourtani$^{2}$, 
Yao Wang$^3$,
Li Feng$^1$,
\thanks{%
New York University Grossman School of Medicine, %
\{%
$^1$Radiology Department, %
$^2$Neurology Department,%
\},
New York, NY 10016, USA.%
}
\thanks{%
$^3$New York University Tandon School of Engineering, %
Electrical and Computer Engineering Department, %
Brooklyn, NY 11201, USA.%
}
\thanks{Please send all correspondence regarding to this manuscript to N. Janju\v{s}evi\'{c} (email:nikola.janjusevic@nyulangone.org).}
\vspace*{-20pt}}%

\markboth{}%
{Shell \MakeLowercase{\textit{et al.}}: A Sample Article Using IEEEtran.cls for IEEE Journals}

\maketitle

\begin{abstract}
Accelerating magnetic resonance imaging (MRI) remains challenging, particularly under realistic acquisition noise. While diffusion models have recently shown promise for reconstructing undersampled MRI data, many approaches lack an explicit link to the underlying MRI physics, and their parameters are sensitive to measurement noise, limiting their reliability in practice. We introduce \emph{Implicit-MAP (ImMAP)}, a diffusion-based reconstruction framework that integrates the acquisition noise model directly into a maximum a posteriori (MAP) formulation. Specifically, we build on the stochastic ascent method of Kadkhodaie et al. and generalize it to handle MRI encoding operators and realistic measurement noise. Across both simulated and real noisy datasets, ImMAP consistently outperforms state-of-the-art deep learning (LPDSNet) and diffusion-based (DDS) methods. By clarifying the practical behavior and limitations of diffusion models under realistic noise conditions, ImMAP establishes a more reliable and interpretable baseline for diffusion-based accelerated MRI reconstruction.
\end{abstract}

\begin{IEEEkeywords}
    Denoising diffusion generative model, inverse problems, deep-learning, denoising, MRI
\end{IEEEkeywords}
\bstctlcite{IEEEexample:BSTcontrol}

\section{Introduction and Background}
Magnetic resonance imaging (MRI) is a widely used, non-invasive technique for the clinical diagnosis of conditions affecting many organs. However, a primary limitation of MRI is its slow acquisition speed, which is constrained by fundamental physical principles. To accelerate scanning, parallel imaging methods use multiple receiver coils to acquire undersampled k-space data, thereby reducing scan time and performing a regularized reconstruction at inference~\cite{Rep2Rep2025}. This regularization is traditionally guided by hand-crafted priors, however, recently end-to-end trained deep neural networks have shown state-of-the-art performance~\cite{lpds2025}. 

Another growing body of related literature is zero-shot deep denoising diffusion generative models (DDGMs) \cite{zhu2023denoising, chung2024decomposed, hoDenoisingDiffusionProbabilistic2020, daras2024survey, rozet2024learning, song2023pseudoinverseguided}. These models have emerged as a powerful class of methods for solving image inverse problems. In natural image domains, these models have demonstrated an exceptional ability to produce realistic and high-fidelity reconstructions. Fundamentally, diffusion models are trained to learn the score function, i.e. the gradient of the log-probability of the data distribution, through a denoising process that gradually removes noise from corrupted samples. By reversing this diffusion process, the models can generate samples from the learned data distribution and effectively recover images from incomplete or degraded measurements~\cite{daras2024survey}. In MRI, diffusion-based approaches have shown promising results in achieving high acceleration rates under simulated noise conditions~\cite{chung2024decomposed}. However, DDGMs are typically formulated using stochastic differential equations or Markov chains, both of which are decoupled from the underlying MRI physics and the  applicability of DDGMs to real-world MRI data with measurement noise introduced by the scanner remains largely unexplored.

In a series of recent publications~\cite{janjusevicCDLNet2022,
janjusevicGDLNet2022, janjusevic2024groupcdl}, we have introduced a framework for the interpretable design of deep denoisers that generalize
effectively across, and even beyond, their training noise levels has been introduced. We further extended this framework to the problem of MRI reconstruction,
demonstrating its applicability in both supervised and self-supervised learning
settings~\cite{lpds2025}. In this work, we expand these ideas to diffusion-based models and
propose an interpretable construction framework for diffusion models tailored to
MRI reconstruction. Specifically, we expand upon the formulation of
diffusion models introduced in Kadkhodaie et al.~\cite{kadkhodaie2021stochastic}
to address inverse problems in the presence of observation noise. We show
state-of-the-art performance of the proposed approach on both synthetically
added noise and real MRI acquisitions with scanner noise.

\subsection{MRI Reconstruction Problem}
We model our observations as an undersampled multicoil Fourier domain signal ${\y \in \C^{N_sC}}$, with $C$ coils, of a ground-truth image ${\x \in \C^N}$,
with Fourier sampling locations denoted by index-set $\Omega$ and coil-sensitivity maps ${\bs \in \C^{NC}}$, 
\begin{equation}
\y_c = \IDMAT_{\Omega}\bF( \bs_c \circ \x ) + \bnu_c, \quad \forall~c=1,\dots,C,
\end{equation}
where $\bF$ is the $N$-pixel 2D-DFT matrix, ${\bs_c \in \C^N ~\forall~ c}$, $\circ$ denotes element-wise multiplication, and
${\IDMAT_{\Omega} \in \{0,1\}^{N_s\times N}}$ is the row-removed identity matrix
with kept rows indicated by $\Omega$. We consider our
observations to be noise-whitened such that ${\bnu[n] \sim \N(0,
\bSigma_y)}$ is Gaussian noise with noise-covariance matrix ${\bSigma_y \in
\C^{C\times C}}$ identically distributed for all pixels ${n \in [1,N_s]}$. We
denote this observation model through
the encoding operator $\bA$ defined such that ${\y = \bA\x + \bnu}$. We
say the observation is accelerated by a factor of ${R={N/N_s}}$.
For the remainder of this manuscript, we follow the mathematical notation of~\cite{Rep2Rep2025}.

\subsection{Stochastic Ascent with a Denoiser's Implicit Prior}
In this section, we present the stochastic gradient descent algorithm with a
denoiser's implicit prior of Kadkhodaie et al.~\cite{kadkhodaie2021stochastic}.
Let $\z = \x + \n$ where $\x \sim p$ and $\n \sim g_\sigma = \N(0, \sigma^2\IDMAT)$. 
Then we may obtain the following relation (Tweedie's Formula~\cite{kadkhodaie2021stochastic}) directly from expressing the noisy image distribution $p_\sigma$ 
as the convolution of the noiseless image distribution $p$ and the Gaussian PDF $g_\sigma$,
\begin{align} \label{eq-tweedie-derivation}
    p_\sigma(\z) &= \int_{\R^N} g_\sigma(\z - \x) p(\x) d\x \\
    \sigma^2 \nabla_{\z} \log p_\sigma(\z) &=  \E[\x | \z] - \z.
\end{align}
Now let $f$ be an MMSE Gaussian-noise denoiser, ex. we trained a learned denoiser 
${f = \argmin_f \E[ \norm{\x - f(\z;\sigma)}^2_2 ]}$ across a range of noise-levels $\sigma$. 
Then ${f(\z;\sigma) \approx \E[\x|\z]}$ and we have approximate access to the gradient of the log-probability of the noisy image distribution,
\begin{equation} \label{eq-tweedie-appx}
    \sigma^2\nabla_{\z}\log p_\sigma(\z) = \E[\x|\z] - \z \approx f(\z;\sigma) - \z. 
\end{equation}
Tweedie's formula \eqref{eq-tweedie-appx} tells us that
the MMSE estimate of the ground-truth $\x$ given a noisy $\z$ is achieved by a
single gradient ascent step on the log-probability of $p_\sigma$, ${\E[\x|\z] =
\z + \sigma^2 \nabla_{\z}\log p_\sigma(\z)}$ with step-size $\sigma^2$. However,
$p_\sigma$ is only a Gaussian-blurred version of our true image distribution
$p$. {\it The key-insight of Kadkhodaie et al.~\cite{kadkhodaie2021stochastic} is
that we may interpret the score function of the noisy distribution,
${\sigma^2 \nabla_{\z}\log p_\sigma(\z)}$, as an approximate ascent direction
on the true image distribution}. If we take a fractional step in this
direction, our new estimate will be closer to the true image distribution, less
noisy ($\sigma \downarrow$), and we will have access to an even more accurate
ascent direction w.r.t. the true image distribution. 
Upon this insight, Kadkhodaie et al.~\cite{kadkhodaie2021stochastic} propose the 
following algorithm, which may be interpreted as a coarse-to-fine gradient ascent 
on the log-probability,
\begin{equation} \label{eq-kadkhodaie}
    \z_{t+1} = \z_t + h_t (f(\z_t) - \z_t) + \bepsilon_t,
\end{equation}
with starting point ${\z_0 \sim \N(0, \IDMAT)}$, step-size ${h_t \in
[0,1]}$ and stochastic noise term ${\bepsilon_t \sim \N(0, \gamma_t^2\IDMAT)}$
introduced to help escape local maxima. 
In the same work,
Kadkhodaie et al.~\cite{kadkhodaie2021stochastic} extended their algorithm to
handle conditioning on measurement data, ${\y = \bA\x}$, by performing
maximization of the log-posterior using Baye's rule, i.e, ${\argmax_{\x} \log
p(\x|\y) = \argmax_{\x} \log p(\x) + \log p(\y|\x)}$. However, their method
relied on $\y$ being noise-free and having an easily SVD decomposable
observation operator $\bA$, limiting its direct application to large-scale MR
imaging inverse-problems. 

In light of these limitations, in Section \ref{sec-method} we propose an
extension of Kadkhodaie's measurement-conditioned stochastic ascent algorithm
which accounts for measurement noise and the large-scale nature of the MRI
encoding operator.

\section{Proposed Method} \label{sec-method}
\subsection{The Implicit MAP Algorithm}
Let $\y = \bA\x + \bnu$ where $\bnu \sim \N(0, \bSigma_y)$.
We seek the to reconstruct $\x$ from $\y$ via obtaining a MAP estimate, 
\begin{align}
    \x^{\diamond} = \argmax_{\x}~& p(\x|\y) \\
    = \argmax_{\x}~& \log p(\x) + \log p(\y|\x).
\end{align}
We tackle this problem with a coarse-to-fine stochastic gradient ascent on the log-prior and log-likelihood functions,
\begin{align}
    \z_{t+1} &= \z_t + h_t \sigma_t^2 \left( \nabla_{\z_t} \log p_{\sigma_t}(\z_t) + \nabla_{\z_t} \log p(\y | \z_t) \right) + \bepsilon_t,
\end{align}
using the prior implicitly encoded by a trained denoiser $f$ to access the prior score function via Tweedie's formula \eqref{eq-tweedie-appx}. 
We switch to the notation to variable $\z$ to indicate that our intermediate variables $\z_t$ belong to the noisy image distribution, ${\z_t \sim p_{\sigma_t}}$.
To access the likelihood score-function, ${\nabla_{\z_t}\log p(\y|\z_t)}$, we model $\z_t$ as a noisy version of our desired ground-truth
signal, ${\z_t \sim \N(\x,~\sigma_t^2 \IDMAT)}$.
Following~\cite{song2023pseudoinverseguided, daras2024survey}, we approximate the distribution of a denoised
$\z_t$, i.e. $f(\z_t)$, with a Gaussian distribution (i.e. a Laplace Approximation), $$f(\z_t) \sim
\N(\E[\x|\z_t], \, \bV_t),$$ where the variance-matrix $\bV_t$ is a modeling parameter which we set to ${\bV_t = \frac{\sigma_t^2}{1+\sigma_t^2}\IDMAT}$ in practice, following \cite{song2023pseudoinverseguided}.
We derive the likelihood function as, $${p(\y|\z_t) = \N(\y | \, \bA\E[\x|\z_t], \, \bSigma_t)},$$ 
where ${\bSigma_t = \bSigma_y + \bA\bV_t\bA^H}$.
Taking the gradient of the log-likelihood yields,
\begin{equation}
\begin{aligned} 
    \nabla_{\z_t} \log p(\y | \z_t) &= - \nabla_{\z_t} \tfrac{1}{2}\norm{\bSigma_t^{-1/2}\left(\bA \E[\x|\z_t] - \y\right)}_2^2 \\
        &= - \nabla_{\z_t}\E[\x|\z_t]^H \bA^H \bSigma_t^{-1} \left(\bA \E[\x|\z_t] - \y\right). \label{eq-grad_likelihood}
\end{aligned}
\end{equation}
With $\E[\x|\z_t] \approx f(\z_t)$, we implement $\nabla_{\z_t}\E[\x |
\z_t]^H\bv \approx \bJ^H_t\bv$ with a vector-jacobian product (VJP) with respect to
our trained denoiser $f$\footnote{The VJP operator is accessible via standard automatic differentiation libraries, often via a \texttt{pullback} function.}. This allows us to solve the symmetric linear
system in the likelihood term \eqref{eq-grad_likelihood} via the conjugate
gradient method \cite{song2023pseudoinverseguided}.
Lastly, we employ a stochastic noise term ($\bepsilon_t$) with stochasticity term ${\beta \in (0,1]}$ for purposes of escaping local
minima, achieving the stochastic MAP estimation algorithm with implicit denoiser prior (ImMAP), shown in Algorithm \ref{alg-ImMAP}.

\newcommand{\commentstyle}[1]{\textcolor{gray}{\ttfamily #1}}
\SetCommentSty{commentstyle}

\begin{algorithm2e}[htb]
\caption{ImMAP}
\label{alg-ImMAP}
\DontPrintSemicolon
\SetKwComment{Comment}{}{}
\KwIn{$\y \sim \CN(\bA\x, \, \bSigma_y)$, ~$\bSigma_y$, ~$\bA$, ~$f$,~$\beta=0.01$, ~$\sigma_L=0.01$, ~$h_0=0.01$}

\KwOut{$\x^\diamond = \argmax_{\x} \log p(\x) + \log p(\y | \x)$}

Draw $\z_0 \sim \CN(0, \IDMAT)$\;
$t = 1$, $\sigma_t^2 = 1$\; 

\While{$\sigma_t > \sigma_L$}{
    \tcp{Denoising step and VJP Operator}
    $\hat{\z}_t, \bJ_t^H \gets \mathrm{pullback}(f(\cdot, \sigma_t), \z_t)$\;
    
    $\sigma_t^2 \gets \norm{\hat{\z}_t - \z_t}_2^2 / N$\label{alg-ImMAP-nl-est}\; 
    
    \tcp{$\Pi$GDM \cite{song2023pseudoinverseguided} Laplace Approximation} 
    $\bSigma_t \gets \bSigma_y + \frac{\sigma_t^2}{1+\sigma_t^2} \bA\bA^H$\;

    $\bv_t \gets \bSigma_t^{-1} \left(\bA \hat{\z}_t - \y\right)$ \tcp{solve via cg}
    
    $\bu_t \gets - \mathbf{J}_t^H \bA^H \bv_t$\;

    \tcp{Stochastic Gradient Ascent}
    $h_t \gets h_0 \frac{t}{1+ h_0(t-1)}$\;
    
    $\gamma_t^2 \gets \sigma_t^2((1-\beta h_t)^2 - (1-h_t)^2)$\;
    $\bepsilon_t \sim \CN(0,\gamma_t^2\IDMAT)$\;

    $\z_{t+1} \gets \z_t + h_t \left(\hat{\z}_t - \z_t + \sigma_t^2\bu_t \right) + \bepsilon_t$\;
    
    $t \gets t + 1$\;
}
    \Return{$\x^\diamond = \z_t$}
\end{algorithm2e}

\subsection{Comparison of ImMAP to Other Diffusion-based Methods}
Denoising diffusion generative
models (DDGMs) are formulated by modeling the gradual transition of a noise-free image
distribution $p_0$ into a distribution of pure noise $p_T$, either through a
Markov-chain process or more generally as a continuous time stochastic
differential equation with forward (corruption) process,
\begin{equation} \label{sde-forward}
    d\x_t = h(\x_t, t) d t + l(t) d\bW_t,
\end{equation}
with $\x_0 \sim p_0$ and $\bW_t$ denoting a Wiener process. 
Solving \eqref{sde-forward} backwards in time is then interpreted as obtaining a sample from the
noise-free image distribution $p_0$, for appropriately chosen $h$ and $l$~\cite{daras2024survey}.
These equations can be easily extended to sampling from a conditional distribution $p_0(\x|\y)$ 
given measurements $\y$ \cite{daras2024survey}. 

This formulation of DDGMs introduces a plethora of modeling choices/assumptions having
solely to do with the SDE/Markov process, unrelated to any imaging inverse problem. In contrast, 
the derivation of ImMAP makes no such modeling assumptions 
and instead relies only on Tweedie's formula \eqref{eq-tweedie-appx}~\cite{daras2024survey} and MRI physics.

Recently, Chung et al. proposed Decomposed Diffusion Sampling
(DDS)~\cite{chung2024decomposed}, a DDGM designed specifically for large scale
noisy inverse problems. DDS is formulated as a Krylov-subspace method, relying
on a ``few-step" conjugate gradient algorithm to keep its intermediate estimates
close to a Krylov-subspace tangent to the noisy image manifold. However, the
existence of such a tangent space is not necessarily valid in practice, and
their choice of number of conjugate-gradient steps is shown to greatly influence
their algorithm's performance~\cite{chung2024decomposed}. Additionally, they introduce another
hyperparameter $\gamma$ to control the trade-off between data-consistency and
their score-based denoiser, which we find in practice needs tuning for different MRI noise-levels~(see Section \ref{sec-exp-ablation}).

\section{Results}
\begin{figure*}[tbh]
    \centering
    \includegraphics[width=0.9\textwidth]{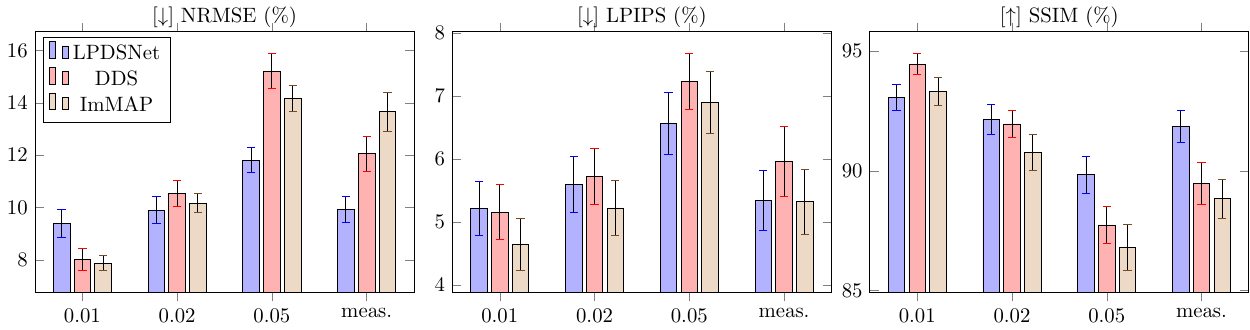}
    \caption{
        Quantitative performance of end-to-end (LPDSNet),
        and denoising diffusion based (DDS, ImMAP) reconstruction methods on
        the T1w brain test dataset. NRMSE/LPIPS/SSIM (\%) shown at $8\times$
        acceleration and across noise-levels $(\sigma_y)$.
        Noise is considered in two scenarios: simulated noise added
        to the coil-combined fully-sampled image data (i.e.
        ${(\bSigma_y)_{cc} = \sigma_y \in (0.01,0.02,0.05)}$), and measurement noise
        (meas., $\bSigma_y=\hat{\bSigma}_y$) where the original fully-sampled
        k-space data is retrospectively undersampled. Error bars show
        standard-error of the mean over 16 test-volumes. 
    }
    \label{fig-tablebar}
\end{figure*}
\begin{figure*}[tbh]
    \centering
    \includegraphics[width=1.5\columnwidth]{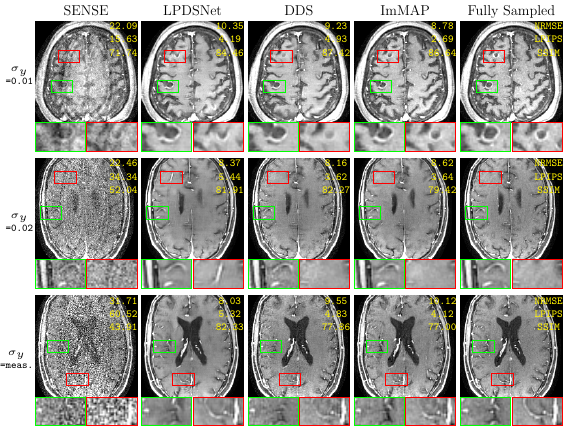}
    \caption{
        Visual comparison of ImMAP joint denoising and reconstruction MRI at
        $8\times$ acceleration against classical reconstruction (SENSE~\cite{espirit}),
        end-to-end deep reconstruction (LPDSNet~\cite{lpds2025}), denoising diffusion model
        (DDS~\cite{chung2024decomposed}), across noise-levels $(\sigma_y)$. Quantitative metrics
        (NRMSE/LPIPS/SSIM (\%)) are shown in yellow above each reconstruction. 
    }
    \label{fig-visuals}
\end{figure*}
\begin{figure}[tbh]
    \centering
    \includegraphics[width=\columnwidth]{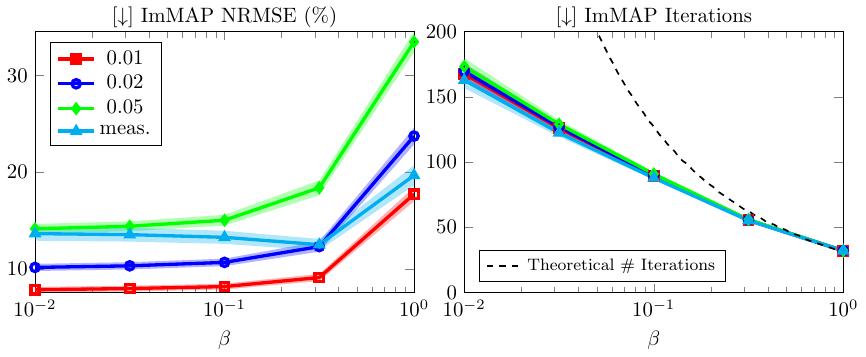}
    \caption{%
        Ablation of ImMAP's noise-injection parameter $\beta \in (0, 1]$. {\bf Left}:
        average NRMSE over the T1w brain test set vs. $\beta$, across noise-levels
        ${\sigma_y=(0.01,0.02,0.05, \text{meas.})}$. Standard error of the mean
        shown via shaded regions. {\bf Right}: Corresponding average number of
        iterations vs. $\beta$, across noise-levels. 
    }
    \label{fig-immap-plot}
\end{figure}
\begin{figure}[thb]
    \centering
    \includegraphics[width=0.55\columnwidth]{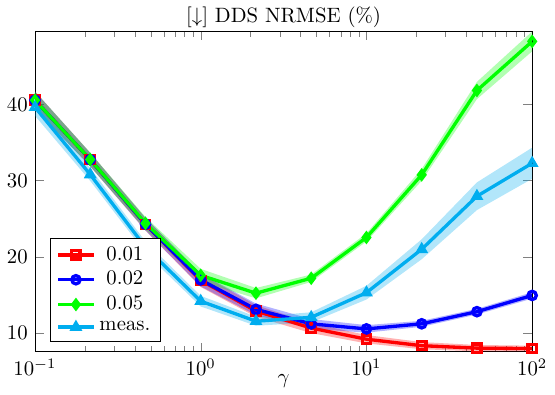}
    \caption{%
        DDS performance (average volume NRMSE) on $8\times$ accelerated T1w brain
        test-set with varying $\gamma$ hyperparameter over different noise
        levels ${\sigma_y=(0.01,0.02,0.05, \text{meas.})}$. Standard error of
        the mean shown via shaded regions. DDS takes 83 iterations.
    } 
    \label{fig-ablation-a}
\end{figure}
\begin{figure}[thb]
    \centering 
    \includegraphics[width=\columnwidth]{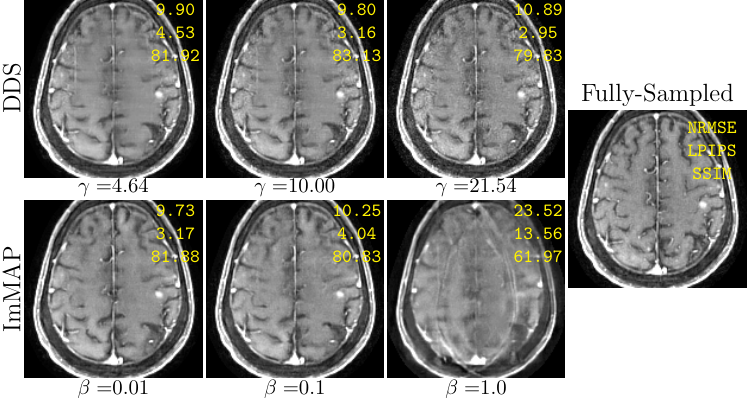}
    \caption{
        Visualization of $8\times$ reconstruction at $\sigma_y=0.02$ when
        changing hyperparameters for DDS ($\gamma$) and ImMAP ($\beta$) on a
        slice from the T1w brain test-set. NRMSE/LPIPS/SSIM $(\%)$ shown on each image as compared to the fully-sampled reconstruction.
    }
    \label{fig-ablation-b}
\end{figure}

\subsection{Dataset and Experimental Settings}
The performance of all models was evaluated on an internal MRI brain dataset
consisting of fully-sampled T1w multicoil k-space data (i.e. ${\y = \bF\bS\x +
\bnu}$). 66 fully-sampled volumes ($\sim$60 slices each) were used
for training with 16 used as a held-out test set. This data was used in two
experimental settings,
\begin{enumerate}
    \item {\bf simulated noise}: the
        {\it fully-sampled coil-combined image} reconstruction
        (${\x^\prime = \bS^H\bF^H\y}$) is passed
        through an MRI encoding operator ${\bA = \IDMAT_{\Omega}\bF\bS}$,
        \begin{equation}
            \y^\prime = \bA\x^\prime + \bnu^\prime,
        \end{equation}
        with noise $\bnu^\prime \sim \N(0,\,\sigma_y\IDMAT)$ added in simulation. Reconstruction involves recovering
        ${\x^\prime}$ from k-space $\y^\prime$, which contains measurement coil-noise.
        Quantitative error measurement compares reconstructions $\hat{\x}$
        to $\x^\prime$ with no irreducible error due to noise. \uline{For simulated noise experiments, white Gaussian noise (${\bSigma_y = \diag(\sigma_y)}$) was added under three noise-levels, ${\sigma_y \in (0.01, 0.02, 0.05)}$}.

    \item {\bf measurement noise}: the {\it fully-sampled multi-coil k-space} is retrospectively undersampled,
        \begin{equation}
            \y^\prime = \bI_{\Omega}\y = \bI_{\Omega}(\bF\bS\x + \bnu) = \bA\x + \IDMAT_{\Omega}\bnu,
        \end{equation}
        with noise $\bnu$ coming directly from the scanner. Reconstruction
        involves recovering the ground-truth noise-free image $\x$ from k-space
        $\y^\prime$. Note that $\x$ does not contain measurement noise and is
        inaccessible to us (due to the SNR of the collected k-space data).
        Quantitative error measurement compares reconstructions $\hat{\x}$ to
        ${\x^\prime}$ with irreducible error due to subsampled measurement noise
        ${\IDMAT_{\Omega}\bnu}$, which is inseparable from the measurement
        k-space data (and hence inseparable from $\x^\prime$). For measurement noise experiments, estimated coil-noise
        covariance matrices $\hat{\bSigma}_y$ were obtained via a wavelet-filter
        multicoil image domain method, following \cite{Rep2Rep2025}. \uline{We
        denote experiments in the measurement noise setting as
        ``$\sigma_y=\text{meas.}$'' for notational brevity}.
        
\end{enumerate}
For all experiments, $8\times$ Cartesian subsampling was used in a single
readout direction with a $4\%$ wide region of k-space included. Coil
sensitivity maps were obtained by the ESPIRiT algorithm~\cite{espirit} applied to the center
of k-space.

\subsection{Training and Implementation}
We trained a single noise-adaptive denoiser $f$ (using the LPDSNet architecture~\cite{lpds2025}
with $\bA = \IDMAT$) on fully-sampled reconstructions of the T1w brain
data with synthetic additive Gaussian white noise (AWGN) in the range $\sigma_t \in [0,0.5]$.
This denoiser was used for both the ImMAP and DDS based inferences for fair comparison. 
We trained two noise-adaptive end-to-end reconstruction networks
(LPDSNet~\cite{lpds2025}) on $8\times$ accelerated simulated noise data and
measurement noise data. Both the denoising LPDSNet (used by ImMAP and DDS) and
reconstruction LPDSNet shared the same architecture hyperparameters for fair
comparison (30 layers, 169 channels, and $7\times 7$ conv. kernels).
Hyperparameters of DDS~\cite{chung2024decomposed} were set as $M=5$, $\eta=0.8$, 83 iterations, and $\gamma$ set by performing a grid-search per noise-level (see Section \ref{sec-exp-ablation}).

\subsection{Joint Reconstruction and Denoising Performance}\label{sec-exp-performance}
Fig.\,\ref{fig-tablebar} shows the quantitative performance of ImMAP against
the end-to-end reconstruction network LPDSNet~\cite{lpds2025} (competitive with
E2EVarNet~\cite{sriram2020end}) and the state-of-the-art denoising diffusion
model DDS~\cite{chung2024decomposed}. Performance was measured across three
metrics: Normalized Root-Mean-Squared-Error ([$\downarrow$] NRMSE), Learned
Perceptual Image Patch Similarity (LPIPS)~\cite{zhang2018perceptual}
([$\downarrow$] LPIPS), and the Structural Similarity Image Metric~\cite{ssim}
([$\uparrow$] SSIM). Across all metrics, ImMAP performs competitively with 
DDS. ImMAP is favored by LPIPS over DDS, especially in the low-noise
range (${\sigma_y=0.01}$). The performance on measurement noise the setting
(${\sigma_y=\text{meas.}}$) heavily favors the end-to-end reconstruction of
LPDSNet, likely due to the oversmoothing tendencies of end-to-end
networks~\cite{chung2024decomposed} being favored by the presence of the
irreducible noise term in image quality metric computations.

Fig.\,\ref{fig-visuals} shows representative visual examples from each
reconstruction method across noise-levels, with addition of the classical
reconstruction baseline method SENSE~\cite{espirit}. Quantitative metrics are
included for each reconstruction. We observe that ImMAP has very strong
low-noise (${\sigma_y=0.01}$) performance, recovering lesions missed/obscured by
SENSE, LPDSNet, and DDS (see image zooms).
For the visual on ${\sigma_y=0.02}$, we observe competitive performance across all quantitative metrics, however, visual inspection reveals a stark difference. LPDSNet is shown to fail to remove aliasing artifacts and DDS is shown to produce noise-like artifacts around the brain edges, while ImMAP achieves superior visual quality.
For ${\sigma_y=\text{meas.}}$, we observe that ImMAP is able to recover fine
texture and details while LPDSNet appears to oversmooth image regions and DDS
appears again to introduce noise-like artifacts.

\subsection{Ablation of ImMAP and DDS}\label{sec-exp-ablation}
The proposed ImMAP demonstrates strong reconstruction stability 
w.r.t. its noise-injection/stochasticity parameter ($\beta$), where $\beta \rightarrow 0$ 
achieves maximum noise-injection and $\beta=1$ removes noise-injection. 
As noise injection is introduced to help escape local-maxima, we expect this parameter to 
improve reconstruction performance as $\beta \rightarrow 0$. 
Additionally, $\beta$ implicitly defines a noise-schedule for $\sigma_t$, and hence
the number of iterations of ImMAP until convergence. This schedule is not
followed strictly due to the noise-level estimation of ImMAP (see Line \ref{alg-ImMAP-nl-est} of Alg.\ref{alg-ImMAP}).
Fig.\,\ref{fig-immap-plot} plots the quantitative reconstruction performance and iteration count of ImMAP with varying $\beta$. We see that, across noise-levels, performance of ImMAP improves as $\beta \rightarrow 0$, with
similar reconstruction performance for a large range of values until $\beta > 0.3$. 
This is with the exception of reconstruction in the measurement noise setting,
where irreducible measurement noise is likely favoring a reconstruction with
less fidelity at high $\beta$.
As $\beta$ increases, we also see that the algorithm takes fewer iterations to converge.
As in Kadkhodaie et al.~\cite{kadkhodaie2021stochastic}, we find that the number of iterations 
until convergence is much lower than theoretically determined by the
$\beta$-defined noise-schedule of $\sigma_t$.
It is clear that $\beta > 0$ should be set as low as possible (0.01 is
sufficient), taking into account any possible computational budget.

State-of-the-art diffusion model DDS~\cite{chung2024decomposed} has its own
hyperparameters, of which $\gamma$, corresponding to a Lagrange multiplier term
in its conjugate gradient solves, requires careful tuning.
Fig.\,\ref{fig-ablation-a} shows the performance of DDS (NRMSE) vs. $\gamma$,
across noise-levels.  We observe that the optimal point (minimum) of each
noise-level curve occurs in a different location.  In Section
\ref{sec-exp-performance}, we used the optimal $\gamma$ determined from
these curves.  In practice, despite fully-sampled reference data being used for
training, fully-sampled reference data at the desired inference noise-level may
not be available, making this grid-search for $\gamma$ difficult or
uninformative. In contrast, the optimal choice of hyperparameter ($\beta$) for
ImMAP is much clearer.

Fig.\,\ref{fig-ablation-b} shows a visual comparison of hyperparameter changes
for both DDS and the proposed ImMAP. DDS shows smoothing and artifact
introduction with decreasing $\gamma$. Additionally, the optimal setting of
$\gamma$ on this example reconstruction is conflicting between image metrics.
In contrast, ImMAP shows similar visual quality for both $\beta=0.01$ and $\beta=0.1$, 
with better quantitative performance across all metrics at lower $\beta$.

\section{Conclusion}
We introduced ImMAP, a diffusion-based MRI reconstruction method that
integrates acquisition physics directly into a MAP estimation framework.
Experiments on both simulated noise and measurement noise reconstruction tasks
demonstrate that ImMAP outperforms state-of-the-art deep learning and
diffusion-based approaches, preserving anatomical detail without noise
amplification or oversmoothing. Notably, it achieves robust performance
without the parameter tuning/sensitivity of existing diffusion methods.
Together, these results establish ImMAP as a reliable and interpretable
baseline, clarifying the practical utility and limitations of diffusion models
for accelerated MRI reconstruction.

\section*{Acknowledgments}
This work was supported in part by the NIH (R01EB030549, EB031083, R21EB032917,
and P41EB017183) and was performed under the rubric of the
Center for Advanced Imaging Innovation and Research (CAI2R), an NIBIB National
Center for Biomedical Imaging and Bioengineering.

\clearpage
\ifCLASSOPTIONcaptionsoff%
  \newpage
\fi
\bibliographystyle{IEEEtran}
\bibliography{IEEEabrv,references}

\end{document}